\documentclass[twocolumn,tighten,times]{aastex62}
\usepackage{color}
\usepackage{multirow}
\usepackage{chngcntr}
\usepackage{mathtools}
\usepackage{amsmath}
\usepackage{amsfonts}
\usepackage{amssymb}
\usepackage[utf8]{inputenc}  
\usepackage{graphicx}        
\usepackage{perpage}         
\usepackage{bm}
\MakePerPage{footnote}

\graphicspath{{./}{figures/}}

\makeatletter

\newcommand{\Rmnum}[1]{\expandafter\@slowromancap\romannumeral #1@}
\makeatother

\newcommand{\Oiii}{[O\,{\sc iii}]}

\newcommand{\Sii}{[S\,{\sc ii}]}
\newcommand{\Nii}{[N\,{\sc ii}]}

\newcommand{\Oabund}{12+$\log$(O/H)}

\shorttitle{Spectroscopy for a UDG with bulge}
\shortauthors{Rong et al.}

\begin{document}

\title{The Internal Kinematics, Stellar Population, and Gas-phase Properties of The Pseudobulge in An Ultra-diffuse Galaxy: AGC721966}

\correspondingauthor{Yu Rong}
\email{rongyua@ustc.edu.cn}

\author{Shihong Liu}
\affiliation{Department of Astronomy, University of Science and Technology of China, Hefei, Anhui 230026, China}
\affiliation{School of Astronomy and Space Sciences, University of Science and Technology of China, Hefei 230026, Anhui, China}

\author{Yu Rong$^*$}
\affiliation{Department of Astronomy, University of Science and Technology of China, Hefei, Anhui 230026, China}
\affiliation{School of Astronomy and Space Sciences, University of Science and Technology of China, Hefei 230026, Anhui, China}

\author{Huiyuan Wang}
\affiliation{Department of Astronomy, University of Science and Technology of China, Hefei, Anhui 230026, China}
\affiliation{School of Astronomy and Space Sciences, University of Science and Technology of China, Hefei 230026, Anhui, China}

\author{Hong-Xin Zhang}
\affiliation{Department of Astronomy, University of Science and Technology of China, Hefei, Anhui 230026, China}
\affiliation{School of Astronomy and Space Sciences, University of Science and Technology of China, Hefei 230026, Anhui, China}

\author{Tie Li}
\affiliation{Department of Astronomy, University of Science and Technology of China, Hefei, Anhui 230026, China}
\affiliation{School of Astronomy and Space Sciences, University of Science and Technology of China, Hefei 230026, Anhui, China}

\author{Yao Yao}
\affiliation{Department of Astronomy, University of Science and Technology of China, Hefei, Anhui 230026, China}
\affiliation{School of Astronomy and Space Sciences, University of Science and Technology of China, Hefei 230026, Anhui, China}

\author{Zhicheng He}
\affiliation{Department of Astronomy, University of Science and Technology of China, Hefei, Anhui 230026, China}
\affiliation{School of Astronomy and Space Sciences, University of Science and Technology of China, Hefei 230026, Anhui, China}

\author{Teng Liu}
\affiliation{Department of Astronomy, University of Science and Technology of China, Hefei, Anhui 230026, China}
\affiliation{School of Astronomy and Space Sciences, University of Science and Technology of China, Hefei 230026, Anhui, China}




\author{Enci Wang}
\affiliation{Department of Astronomy, University of Science and Technology of China, Hefei, Anhui 230026, China}
\affiliation{School of Astronomy and Space Sciences, University of Science and Technology of China, Hefei 230026, Anhui, China}

\author{Cheng Cheng}
\affiliation{National Astronomical Observatories, Chinese Academy of Sciences, Beijing 100012, China}

\author{Xu Kong}
\affiliation{Department of Astronomy, University of Science and Technology of China, Hefei, Anhui 230026, China}
\affiliation{School of Astronomy and Space Sciences, University of Science and Technology of China, Hefei 230026, Anhui, China}



\begin{abstract}

	Leveraging spectroscopic data from the Sloan Digital Sky Survey, we conduct a comprehensive analysis of the central stellar velocity dispersion, stellar population properties, star formation history, and gas-phase chemical abundances in AGC721966, a unique ultra-diffuse galaxy (UDG) harboring a pseudobulge. Our findings reveal that the pseudobulge formed in the early universe but underwent a recent episode of rejuvenated star formation. The system exhibits a mass-weighted (light-weighted) stellar population age of $\tau_{\star}\sim 7.4\pm2.5$ ($2.9\pm1.5$)~Gyr, a stellar metallicity of [M/H]$\sim -0.62\pm0.26$ ($-0.55\pm0.20$), an $\alpha$-element enhancement of [$\alpha$/Fe]$\sim 0.36\pm0.09$ ($0.37\pm0.07$), and a gas-phase oxygen abundance of \Oabund$\sim 8.15\pm0.03$. The central stellar velocity dispersion is measured as $\sigma_{\rm c}\sim 57.9\pm15.7$~km/s. These results provide robust evidence supporting the early halo-halo merging formation scenario proposed by \cite{Rong25}, while unequivocally ruling out the ``failed'' $L^{\star}$ formation model, at least for AGC721966. Furthermore, through systematic application of the baryonic Tully-Fisher relation, we establish that these pseudobulge-hosting UDGs are neither misidentified nuclear star cluster-bearing dwarf galaxies nor bulge-dominated massive galaxies, thereby affirming their distinct evolutionary pathway.

\end{abstract}


\section{Introduction} \label{sec:1}

Ultra-diffuse galaxies \citep[UDG;][]{vanDokkum15} are characterized by effective radii comparable to the Milky Way yet stellar masses akin to dwarf galaxies. Early observations of UDGs in galaxy clusters suggested a high dark matter fraction, inferred from their resilience to tidal disruption, which posited that diffuse stellar distributions might be shielded by massive dark matter halos. The detection of a putative $10^{12}\ M_{\odot}$ \citep{vanDokkum16} halo in the Coma UDG, DF44, initially bolstered this hypothesis \citep{vanDokkum16}, prompting speculation that UDGs represent ``failed L* galaxies'' \citep[FLGs;][]{vanDokkum15}\---systems stunted in star formation efficiency by early environmental effects or internal feedback \citep{Yozin15}. However, subsequent studies leveraging weak lensing, globular cluster counts, and X-ray data revealed that most UDGs reside in dwarf-scale halos \citep{Sifon18,Amorisco18,Marleau24,Kovacs20}, though a subset with abundant globular clusters may still align with FLG scenarios \citep{Lim20,Gannon22}.

A paradigm shift occurred with the discovery of UDGs DF2 and DF4 \cite{vanDokkum18,vanDokkum19}, which exhibit negligible dark matter content. These findings challenge empirical galaxy evolution models, which predict halo masses of $10^{10}\--10^{11}\ M_{\odot}$ for galaxies with UDG-like stellar masses ($10^7\--10^9\ M_{\odot}$). Such discrepancies\---whether excess or deficit\---underscore that UDG formation mechanisms may fundamentally diverge from canonical pathways.

Numerous models attempt to reconcile these observations. High-spin halos \citep{Rong17a,Amorisco16,Liao19,Benavides23,Rong24a,Rong20c} and supernova-driven outflows \citep{DiCintio17,Chan18,Cardona-Barrero20} reproduce UDG scaling relations but struggle with overproduction or structural inconsistencies. Alternative frameworks invoke dwarf mergers \citep{Wrigts21}, tidal heating \citep{Carleton19,Jiang19}, ram-pressure stripping \citep{Grishin21}, or self-interacting dark matter \citep{Zhang25}.

Recent observations of field UDGs with pseudobulges \citep{Rong25}\---unprecedented in dwarf galaxies \citep{Kormendy10,Kormendy12}\---offer a new avenue to probe their origins.  These systems host extended, rotationally supported disks alongside pseudobulges (S\'ersic index $n<2.5$, stellar mass $M_{\star}\sim 10^{8.0}\--10^{8.7}\ M_{\odot}$, and effective radius $r_{\rm{h}}\sim 300-700$~pc), far exceeding nuclear star cluster dimensions \citep{Walcher06,Georgiev16,Spengler17,Neumayer20,Hilker99,Drinkwater00}. Their elevated HI-derived rotation velocities and pseudobulge presence suggest halo masses exceeding typical dwarfs, potentially bridging UDGs to FLGs or early gas-rich mergers \citep{Bekki08}.

Resolving whether these UDGs harbor supermassive black holes (SMBHs) or originate from halo mergers requires empirical validation of their star formation histories (SFHs), stellar kinematics, and halo masses. Here, we present spectroscopic analysis of the pseudobulge-hosting UDG AGC721966, quantifying its central velocity dispersion, stellar population parameters, and gas-phase abundances. By coupling these results with dynamical mass estimates and baryonic Tully-Fisher constraints, we probe the system's formation pathway and dark matter content.
The paper is structured as follows: Section~\ref{sec:2} details spectral fitting techniques and derives the central velocity, stellar population, gas-phase chemical abundance; Section~\ref{sec:3} interprets the derived black hole mass and halo mass, evaluates formation scenarios, and recalibrates distances via baryonic Tully-Fisher relations. Conclusions are synthesized in Section~\ref{sec:4}. Throughout this paper, ``$\log$'' represents ``$\log_{10}$''.

\section{Spectroscopy for the pseudobulge of AGC721966}\label{sec:2}

\begin{figure*}[!]
\centering
\includegraphics[angle=0,width=0.99\textwidth]{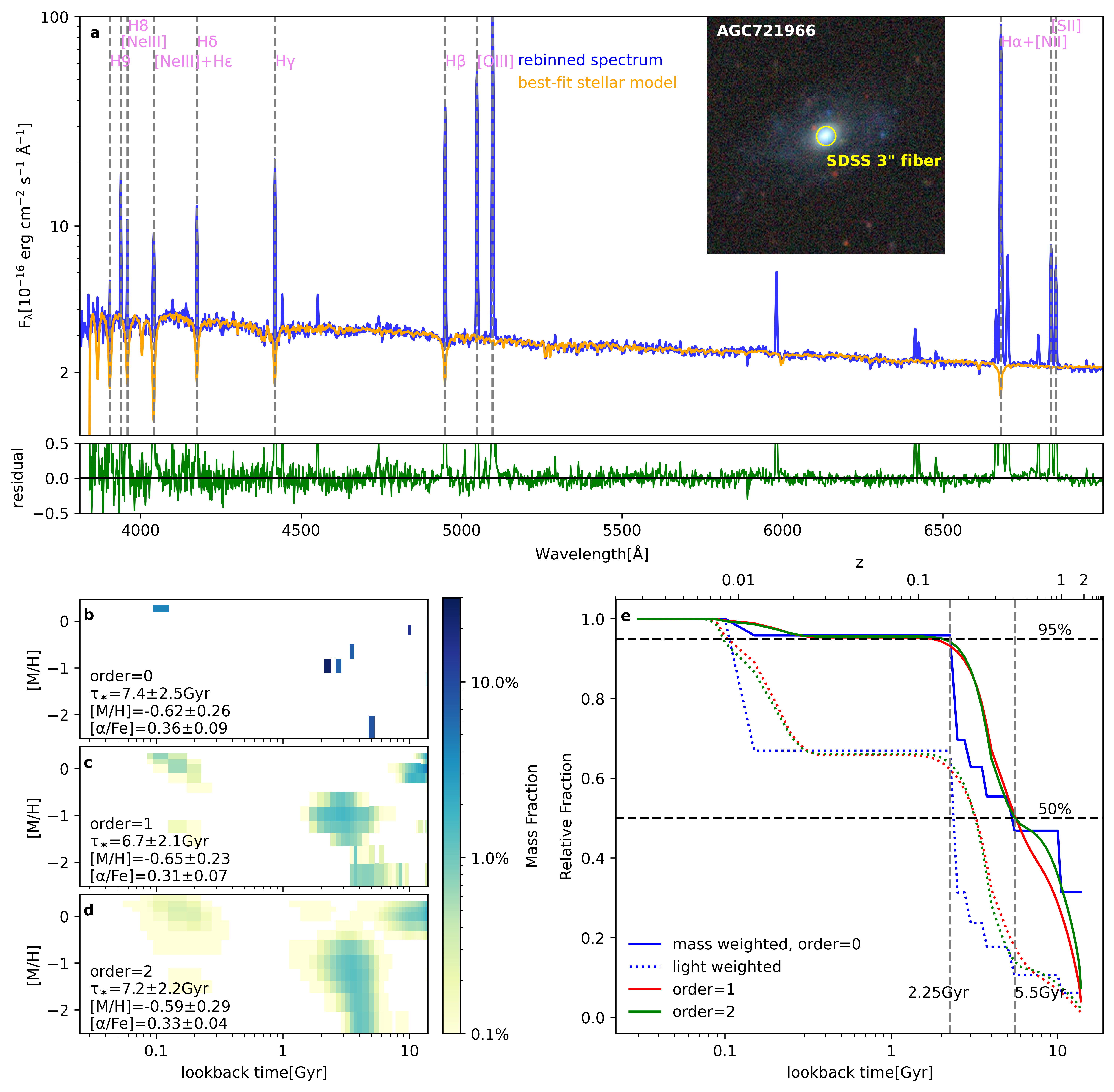}
\caption{Stellar population analysis of the pseudobulge in AGC721966. Panel~a presents the SDSS spectrum of the pseudobulge after rebinning (blue), best-fit stellar population model (orange), and residuals (green). The inset displays the image of AGC721966, highlighting the SDSS 3~arcsec spectroscopic fiber positioned on the pseudobulge of the galaxy. Panels~b, c, and d show  the mass-weighted stellar age-metallicity distributions, illustrating the SFH derived without regularization (b), with first-order regularization (c), and with second-order regularization (d). The color bar indicates the mass fractions corresponding to each combination of age and metallicity. Panel~e: mass assembly (solid lines) and light assembly (dotted lines) of the pseudobulge, with results shown for the unregularized case (blue), first-order regularization (red), and second-order regularization (green). The horizontal black dashed lines mark the epochs when 50\% and 95\% of the stellar mass or light had been assembled, respectively.}
\label{fitting_result}
\end{figure*}

Among the five pseudobulge-hosting UDGs identified by \cite{Rong25}, AGC721966 stands out as the nearest system, with a distance of $78.9\pm2.3$~Mpc. Spectroscopic data for this galaxy were obtained from the Sloan Digital Sky Survey \citep[SDSS; Data Release 12 (DR12),][]{Alam15}, targeting its central region through a $3''$-diameter fiber aperture, as shown by the insert of panel~a in Fig.~\ref{fitting_result}. At the distance of AGC721966, this aperture corresponds to a physical scale of approximately 1.1~kpc, fully encompassing the pseudobulge (effective radii $r_{\rm h}\sim 0.4$~kpc in $g$-band and 0.60~kpc in $r$-band, respectively). The central surface brightness of the pseudobulge in the $g$-band ($\mu_g\sim 16\ \rm mag/arcsec^2$) starkly contrasts with that of the surrounding disk ($\mu_g\sim 25\ \rm mag/arcsec^2$), indicating that the disk contributes negligibly to the central luminosity ($<0.1\%$ of the total flux within the fiber). Consequently, the SDSS spectrum predominantly reflects the pseudobulge's properties, with negligible contamination from the disk.

The spectrum exhibits a signal-to-noise ratio (S/N) of $\sim 25$ (measured across 5450\--5550~\AA), sufficient for robust stellar dynamical and population synthesis analyses following spectral rebinning. Prior to analysis, the spectrum was corrected for Galactic extinction using the \cite{Fitzpatrick99} extinction law with $R_V=3.1$ and $E(B-V)$ values retrieved from the NASA/IPAC Extragalactic Database (NED).


\subsection{Stellar velocity dispersion}\label{sec:2.1}

Building upon the methodology of \cite{Rong20}, we employ the penalized pixel-fitting code \textsc{pPXF} \citep[v7.3.0;][]{Cappellari17} to perform spectral fitting and derive the central stellar velocity dispersion $\sigma_{\rm{c}}$ of the pseudobulge in AGC721966. Our implementation adopts 6th-order additive (``degree'') and multiplicative (``mdegree'') polynomials, a configuration optimized to mitigate template mismatch and continuum variations \citep{Emsellem04,Guerou17,Spiniello21,Bidaran20}.

To maximize dispersion measurement fidelity, we utilize the Indo-US stellar template library \citep{Valdes04}, selected for its high spectral resolution \citep[FWHM=1.35~\AA;][]{Beifiori11}, and mask all emission-line regions during the fitting process. Following instrumental resolution correction (SDSS spectrum FWHM=2.76~\AA, corresponding to a velocity resolution of $\sim 64$~km/s at 5500~\AA), we measure $\sigma_{\rm{c}}\simeq 57.9\pm 15.7$~km/s for the pseudobulge, where uncertainties are derived via bootstrap resampling as per \cite{Fahrion25,Kacharov18}. This dispersion substantially exceeds values characteristic of nuclear star clusters (NSC; $\sigma_{\rm NSC}\sim 10\--40$~km/s) and globular cluster (GC; $\sigma_{\rm GC}\sim 3\--25$~km/s) observed in UDGs \citep[e.g.,][]{Fahrion25,Khim25} and dwarf galaxies \citep[e.g.,][]{Neumayer20,Nguyen17,Seth10}.
 
Robustness tests confirm that $\sigma_{\rm c}$ remains within the $1\sigma$ uncertainty range when varying polynomial orders (``degree''=4, 5, 7, 8, 10; ``mdegree''=5, 8, 10, 12), demonstrating insensitivity to fitting polynomials choices.

\begin{figure*}[!]
\centering
\includegraphics[angle=0,width=0.99\textwidth]{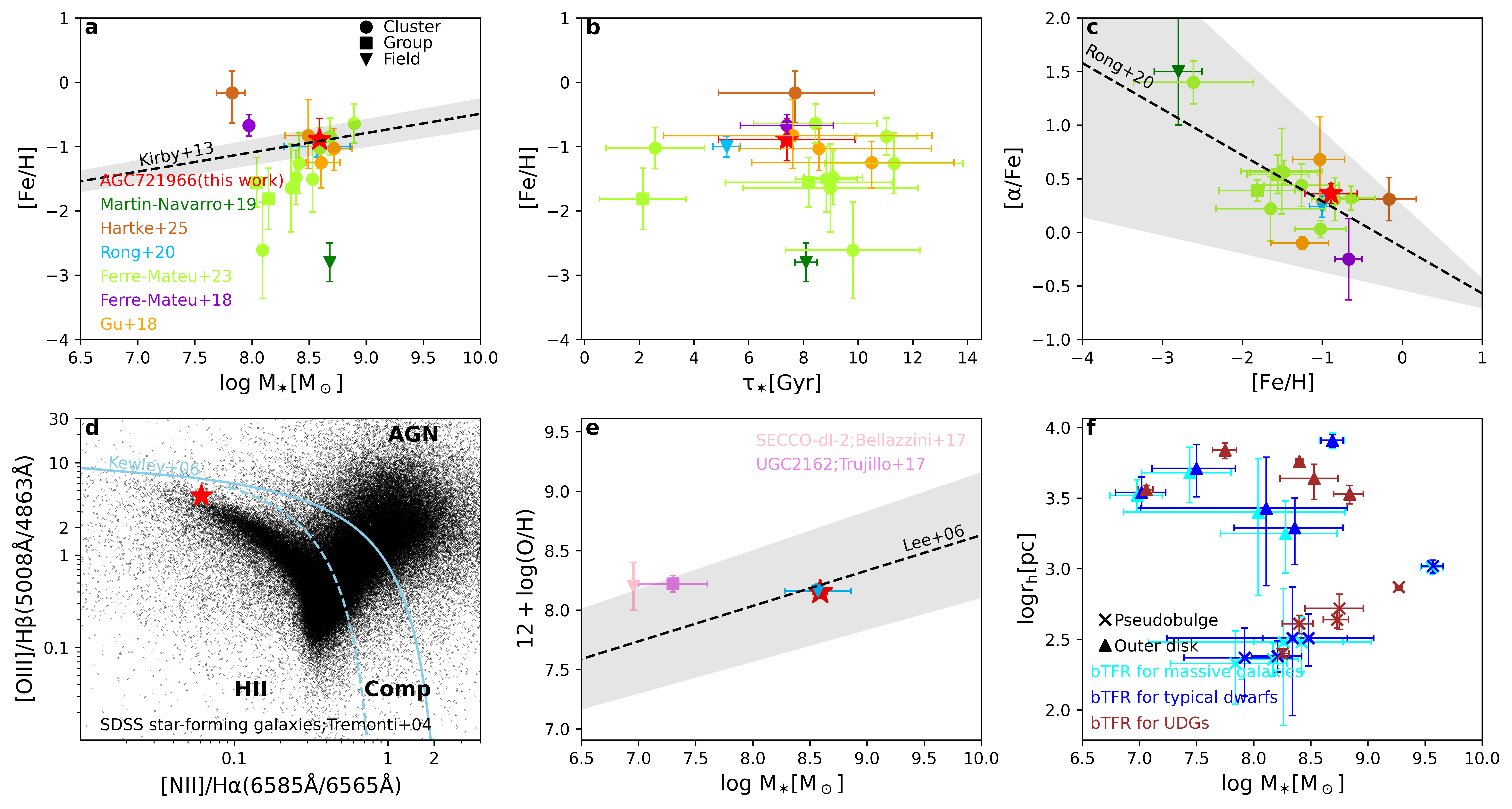}
\caption{Comparative analysis of the pseudobulge in AGC721966 and other UDGs. Symbols denote UDGs in different environments: dots for galaxy clusters, squares for groups, and inverted triangles for field galaxies. The pseudobulge in AGC721966 is highlighted as a red star. Panel~a: the mass-weighted stellar metallicity [Fe/H] as a function of stellar mass $M_{\star}$. The universal [Fe/H]-$M_{\star}$ relation for nearby dwarf galaxies \citep{Kirby13} is shown as a black dashed line, with its $1\sigma$ uncertainty represented by the grey shaded region. Panel~b: [Fe/H] versus mass-weighted stellar ages $\tau_{\star}$. Panel~c: $\alpha$-element enhancement as a function of [Fe/H]. The [$\alpha$/Fe]-[Fe/H] relation for typical UDGs \citep{Rong20} is indicated by a dashed line, with the $1\sigma$ uncertainty region shaded in grey. Note that the results shown in panel~a to c are all mass-weighted. Panel~d: 12+log(O/H) versus $M_{\star}$, with the mass-metallicity relation for nearby star-forming dwarfs \citep{Lee06} overlaid as a dashed line and its $1\sigma$ uncertainty as a grey region. Panel~e: BPT diagram showing SDSS star-forming galaxies (small black dots) and the pseudobulge in AGC721966. The curves for AGN selection \citep{Kewley06} are also displayed.
Panel~f: size-mass relations for pseudobulges (crosses) and outer stellar disks (triangles) of five UDGs hosting pseudobulges \citep{Rong25}, using distances derived from the bTFR. The cyan, blue, and magenta colors correspond to the application of typical bTFRs for gas-dominant massive galaxies \citep{Stark09}, typical dwarf galaxies \citep{Lelli16}, and gas-rich UDGs \citep{Rong24a}, respectively.}
\label{UDG_property}
\end{figure*}

\subsection{Stellar population and SFH}\label{sec:2.2}

To conduct stellar population synthesis (SPS), we employ the MILES single stellar population (SSP) spectral templates \citep{Vazdekis15}, circumventing the absence of age information in the Indo-US library. Although MILES exhibits lower spectral resolution (FWHM$=2.5$~\AA), compared to Indo-US, its comprehensive age-metallicity parameter space renders it optimal for SPS.

Our analysis leverages MILES models calibrated using BaSTI isochrones \citep{Pietrinferni06} and a bimodal Milky Way-like initial mass function (IMF) with a high-mass slope of 1.30. The grid spans 53 ages (from 30~Myr to 14~Gyr) and 12 metallicities (from [M/H]=-2.27 to +0.40). Given the MILES library's restriction to scaled-solar ([$\alpha$/Fe]=0) and $\alpha$-enhanced ([$\alpha$/Fe]=0.4~dex) templates, we adopt the methodology of \cite{Fahrion19} to construct a finely sampled SSP grid. We systematically interpolate between existing templates at 0.1~dex intervals across [$\alpha$/Fe]=0.0 to [$\alpha$/Fe]=0.4~dex, with extrapolation to [$\alpha$/Fe]=-0.1 dex, assuming linear $\alpha$-abundance behavior within this regime. This $\alpha$-variable grid enables robust recovery of average [$\alpha$/Fe] from spectral features.

During the fitting process, we rebin the spectrum by combining every two adjacent data points into a single point to enhance the spectrum SNR. After binning, the spectrum SNR increases to $\sim 33$. We disable additive polynomials and and regularization (``REG\_ORD''=0) in \textsc{pPXF} to minimize overfitting. The rebinned spectrum and best-fit stellar model are shown in panel~a of Fig.~\ref{fitting_result}. The resulting mass-weighted parameters are total-metallicity [M/H]$=-0.62\pm 0.26$, stellar age $\tau_{\star}=7.4\pm2.5$~Gyr, and [$\alpha$/Fe]$\simeq 0.36\pm0.09$; light-weighted values are [M/H]$\sim-0.55\pm0.20$, $\tau_{\star}\sim2.9\pm1.5$~Gyr, and [$\alpha$/Fe]$\sim 0.37\pm0.07$. Iron abundance is derived as [Fe/H]=[M/H]-0.75[$\alpha$/Fe] \citep{Vazdekis15}.

Figure~\ref{fitting_result} compares SSP solutions under first- and second-order regularization. While regularization smoothes age-metallicity template weights\---optimal for systems with prolonged star formation\---all approaches yield parameters consistent within $1\sigma$ uncertainties. Tests with multiplicative polynomial orders (``mdegree''= 5, 8, 10, 12) confirm result stability.

The pseudobulge aligns with the universal [Fe/H]$\--M_*$ relation of dwarf galaxies \citep{Kirby13}, as explored in panel~a of Fig.~\ref{UDG_property}, yet its age mirrors cluster/group UDGs \citep{Gu18,Ferre-Mateu18,Ferre-Mateu23,Hartke25} rather than isolated counterparts \citep{Rong20}, as indicated in panel~b of Fig.~\ref{UDG_property}. Its [$\alpha$/Fe]-[Fe/H] trend matches typical UDGs across environments \citep{Rong20}, as shown in panel~c of Fig.~\ref{UDG_property}. Our results indicate that the pseudobulge of AGC721966 is indeed old, and plausibly formed by starbursts with short star-formation timescales.

As shown in panel~e of Fig.~\ref{fitting_result}, unregularized and regularized SFH reconstructions consistently reveal the early formation of the pseudobulge. Approximately 50\% of the stellar mass of the pseudobulge was formed before lookback time $t\sim 5.5$~Gyr, while 95\% mass was assembled before $t\sim 2.25$~Gyr. This SFH contrasts with the extended star formation in typical UDGs \citep{Ferre-Mateu18,Martin-Navarro19,Rong20} and single-burst SFH of GCs \citep{Holtzman92,OConnell95,Whitmore99}, instead resembling nuclear star clusters in dwarfs \citep{Johnston20,Neumayer20}. Yet it dose not mean that this galaxy is a nucleated dwarf, as discussed in section~\ref{3.4}

\subsection{Emission line properties}

The strong Balmer line emission and a big difference between the mass- and light-weighted ages indicate recent gas accretion and localized star formation rejuvenation.

Following subtraction of the best-fit stellar population model from the stacked spectrum, we perform Gaussian profile fitting to individual emission lines and calculate the H$\alpha$/H$\beta$ flux ratio to estimate dust extinction under Case-B recombination assumptions \citep{Hummer87}. The pseudobulge of AGC721966 occupies the star-forming locus in the Baldwin-Phillips-Terlevich \citep[BPT;][]{Baldwin81} diagnostic diagram, as explored by panel~d of Fig.~\ref{UDG_property}, indicating a low likelihood of active galactic nucleus (AGN) contamination \citep{Kewley01,Kewley06}.

To derive the gas-phase oxygen abundance 12+$\log$(O/H), we implement two robust metallicity diagnostics: the N2S2H$\alpha$ index \citep{Dopita16} and O3N2 calibration \citep{Pettini04}. This approach supersedes traditional [O\,{\sc ii}]/[O\,{\sc iii}]-based methods \citep{Izotov06,Kniazev04}, given the absence of [O\,{\sc ii}]$\lambda3727$ coverage in our spectral window and low SNRs for the [O\,{\sc ii}]$\lambda 7320,7330$ doublet. The N2S2H$\alpha$ diagnostic combines the \Nii$\lambda6584$/H$\alpha$ and \Nii$\lambda6584$/\Sii$\lambda\lambda$6717,31 line ratios, while O3N2 utilizes \Oiii$\lambda5007$/H$\beta$ versus \Nii$\lambda6584$/H$\alpha$.

Both methods yield consistent oxygen abundances for the pseudobulge: \Oabund$\simeq 8.15\pm0.02$ (N2S2H$\alpha$) and $\simeq 8.14\pm0.03$ (O3N2). As shown in panel~e of Fig.~\ref{UDG_property}, this metallicity aligns with the characteristic \Oabund\--$M_*$ relation observed in both gas-rich field UDGs \citep{Rong20,Trujillo17,Bellazzini17} and star-forming dwarf galaxies \citep{Lee06}, reinforcing the pseudobulge's adherence to canonical chemical evolution pathways.


\section{Discussion}\label{sec:3}

\subsection{Supermassive black hole?}\label{sec:bh}

$\sigma_{\rm{c}}$ provides a basis for estimating central supermassive black hole (SMBH) masses via the well-established $M_{\rm{BH}}\--\sigma_{\rm c}$ correlation \citep{Tremaine02,Kormendy13,Greene16}. However, this empirical relation exhibits robust predictive power primarily in elliptical galaxies and systems hosting classical bulges \citep{Haring04,Schutte19}, while remaining less constrained in pseudobulges. Applying the \cite{Tremaine02} relation, $M_{\rm{BH}}=10^{8.13\pm0.06}\ M_{\odot}\ (\sigma_{\rm{c}}/200)^{4.02\pm 0.32}$, to AGC721966 yields $M_{\rm{BH}}\simeq 9.3_{-4.4}^{+6.1}\times 10^5\ \rm M_{\odot}$, placing it near the lower mass boundary of SMBHs  ($M_{\rm{BH}}\gtrsim 10^6\ \rm M_{\odot}$). Given that pseudobulges typically host black hole masses lower by an order of magnitude compared to classical bulges \citep{Kormendy13,Hu08}, our results suggest AGC721966 may lack an SMBH, instead potentially harboring an intermediate-mass black hole (IMBH).

Notably, the measured $\sigma_{\rm{c}}\sim 57.9$~km/s in AGC721966 exceeds the central velocity dispersions of field UDGs \citep[$\sigma_{\star}\sim 35$~km/s;][]{Rong20} by 65\%. This disparity implies that, while still inconsistent with SMBH occupation, the putative IMBH in this system may represent a distinct outlier relative to typical UDGs in similar environments, underscoring the potential presence of rare evolutionary pathways or atypical formation conditions.


\subsection{Massive halo mass?}\label{3.2}

The inferred IMBH in AGC721966 implies a relatively low-mass dark matter halo. Aligning with the empirical stellar-to-halo mass relation \citep[$M_{\star}\--M_{200}$;][]{Girelli20}, we estimate a halo mass of $\log M_{200}\simeq 10.70\pm 0.04$). To independently validate this prediction, we apply an HI kinematics-based approach following \cite{Zhang25} and \cite{Rong24a}, which circumvents assumptions inherent to $M_{\star}\--M_{200}$ or $M_{\rm{BH}}\--M_{200}$ scaling relations.

First, we derive the dynamical mass enclosed within the HI radius $r_{\rm{HI}}$, defined as the radius at which the HI surface density attains $1\ \rm M_{\odot}\rm{pc^{-2}}$, as
\begin{equation}
 M_{\rm dyn} (<r_{\rm HI}) =V_{\rm c}^2 r_{\rm HI}/G\,,
 \label{dm}
\end{equation}
where $G$ is the gravitational constant. $V_{\rm{c}}$ is the rotation velocity \citep{Rong25}. The HI radius is determined via the empirically calibrated relation $\log_{10} r_{\rm{HI}}=0.51\log_{10} M_{\rm{HI}}-3.59$ \citep{Wang16,Gault21}, validated for UDGs \citep{Gault21}.

Second, assuming a Burkert dark matter profile \citep{Burkert95}, we compute the halo mass ($M_{200}$) by solving
 \begin{equation}
\small
\begin{aligned}
&{M_{\rm dyn}(<r_{\rm HI})}-M_{\rm{bar}} =  \int_0^{r_{\rm{HI}}}4\pi r^2\rho_{\rm{B}}(r)\mathrm{d}r \\
 & =  2\pi \rho_{0} r_{0}^3\left[\ln \left(1+\frac{r_{\rm{HI}}}{r_{0}}\right)
 +0.5\ln\left(1+\frac{r_{\rm{HI}}^2}{r_{0}^2}\right) 
 -{\rm arctan}\left(\frac{r_{\rm{HI}}}{r_{0}}\right)\right]\,,
\end{aligned}
  \label{bur}
  \end{equation}
and
\begin{equation}
\small
\begin{aligned}
&M_{200} = \int_0^{R_{200}}4\pi r^2\rho_{\rm B}(r)\mathrm{d}r \\
& =  2\pi \rho_{0} r_{0}^3\left[\ln \left(1+\frac{R_{200}}{r_{0}}\right)+0.5\ln\left(1+\frac{R_{200}^2}{r_{0}^2}\right) 
-{\rm arctan}\left(\frac{R_{200}}{r_{0}}\right)\right]\,,
\end{aligned}
\label{bur_m}
\end{equation}
where $M_{\rm{bar}}\simeq M_{*}+1.33M_{\rm{HI}}$ represents the baryonic mass. $M_{\rm{HI}}$ is the HI mass. $R_{200}$ represents the virial radius enclosing a mean density that is 
200 times the critical value.
The core parameters $r_{0}$ and $\rho_{0}$ are linked via $\log [(r_0/{\rm kpc})] = 0.66-0.58(\log[M_{200}/10^{11} \rm M_{\odot}])$ \citep{Salucci07}.

Halo mass uncertainties are propagated via Monte Carlo simulations ($N=1,000$; realizations), incorporating errors in $M_{\rm{bar}}$ ($\sigma_{M_{\rm{bar}}}=\sqrt{\sigma^2_{M_{*}}+(1.33\sigma_{M_{\rm{HI}}})^2}$), $r_{\rm{HI}}$ ($\sigma_{r_{\rm{HI}}}$), 
and $V_{\rm c}$ ($\sigma_{V_{\rm c}}$), and the $r_0$-$M_{200}$ relation. The final uncertainty combines the standard deviation of simulated masses with the Burkert profile's intrinsic scatter $\sigma'_{M_{200}} = \sqrt{\sigma_{M_{200}}^2 + (0.15~{\rm dex})^2}$ \citep{Wang20}.

This yields $\log M_{200}\simeq 11.14\pm 0.15$, consistent with dwarf-scale halos despite methodological uncertainties \citep[see ][for robustness in HI kinematics for diffuse dwarfs]{Zhang25}.  The HI-derived halo mass exceeds the $M_{\star}\--M_{200}$ prediction by 0.44~dex, yet decisively excludes a massive ($\sim 10^{12}\ M_{\odot}$) dark matter halo for AGC721966.


\subsection{Possible formation mechanism?}\label{3.3}

In our prior investigation of \cite{Rong25}, we proposed two distinct mechanisms to account for the formation of pseudobulge-hosting UDGs, specifically addressing their evolutionary transition from compact bulge-dominated configurations in the early universe to extended disk-like structures at later epochs.

The first posits a ``failed $L^{\star}$'' galaxy scenario, which necessitates a massive dark matter halo ($M_{200}\gtrsim 10^{12}\ M_{\odot}$) and a central SMBH, to stabilize the system. However, this model conflicts with our observational constraints on AGC721966, which exhibits neither evidence of a massive halo nor SMBH activity. The second mechanism invokes early-universe ``star-free'' mergers \citep{Rong25}, wherein gas-rich progenitor halos drive central starbursts to form compact pseudobulges while transferring orbital angular momentum to generate rotationally supported outer disks. Our spectroscopic analysis—revealing an ancient stellar population ($\tau_{\star}\sim 7.4$~Gyr) and pronounced $\alpha$-element enhancement ([$\alpha$/Fe]$\sim 0.36$)\---aligns with this merger-driven framework. However, while theoretically viable, this formation model demands rigorous validation through numerical simulations and spectroscopic studies for more pseudobulges in UDGs.

\subsection{Nuclear star cluster?}\label{3.4}

While the pseudobulge-hosting UDGs identified by \cite{Rong25} exhibit high radial velocities ($v>5400$~km/s corresponding to redshift $z>0.018$), and reside in isolated environments devoid of galaxy groups/clusters, a potential concern still arises: their radial velocities may contain non-negligible peculiar motion components. Such kinematic contamination could artificially inflate distance estimates via Hubble flow assumptions \cite[e.g.,][]{Haynes18}, potentially misclassifying NSCs as pseudobulges due to spatial resolution limitations.

To address this systematically, we employ the baryonic Tully-Fisher relation (bTFR) as an independent distance calibrator. Three distinct bTFR calibrations are applied: (1) the typical bTFR for gas-dominant massive galaxies \citep{Stark09}, (2) the bTFR for typical dwarf galaxies \citep{Lelli16}, as well as (3) the bTFR for gas-rich UDGs \citep{Rong24a}. As demonstrated in panel~f of Fig.~\ref{UDG_property}, all three relations robustly confirm the extended nature of these central structures ($r_{\rm{h}}>200$~pc)\---far exceeding the characteristic sizes of NSCs, UCDs, and GCs which typically have effective radii ranging from approximately 3~pc and 100~pc \citep{Walcher06,Georgiev16,Spengler17,Neumayer20,Hilker99,Drinkwater00}.

Furthermore, the significant ellipticity ($\epsilon=0.28$ in the $r-$band) observed in AGC721966's pseudobulge disfavors NSC interpretations, as nuclear clusters typically exhibit near-circular morphologies, also disfavoring the nuclear star cluster assumption. Collectively, these findings establish that these galaxies, including AGC721966, are very likely to be pseudobulge-hosting UDGs\---a morphological class previously unrecognized in galaxy surveys.


\section{Conclusion}\label{sec:4}

Our analysis robustly excludes the possibility that pseudobulge-hosting UDGs, exemplified by AGC721966, represent massive galaxies or misclassified nucleated dwarfs with spurious distance measurements. These systems are unambiguously identified as dwarf galaxies (UDGs) harboring pseudobulges.

The absence of both a SMBH and a massive dark matter halo in AGC721966 disfavors the ``failed L$^{\star}$ galaxy'' hypothesis.

Spectroscopic diagnostics reveal a pronounced  $\alpha$-element enhancement and an ancient stellar population in the pseudobulge. These findings provide robust evidence for the ``star-free'' merger-driven formation model proposed by \cite{Rong25}, i.e., the pseudobulge emerges from dissipative, gas-rich halo-halo mergers at high redshift, triggering rapid centrally concentrated star formation, while the outer stellar disk results from the recent gas accretion onto the merger remnant, whose elevated spin inherited from the progenitor system's orbital angular momentum, channels infalling material into an extended low-surface-brightness disk.


\acknowledgments

YR acknowledges supports from the NSFC grant 12273037, the CAS Pioneer Hundred Talents Program (Category B), and the USTC Research Funds of the Double First-Class Initiative. HXZ acknowledges support from the NSFC grant 11421303. 
HYW is supported
by the National Natural Science Foundation of China (NSFC, Nos. 12192224) and CAS Project for Young Scientists in Basic Research, Grant No. YSBR-062. This work is supported by the China Manned Space Program with grant no. CMS-CSST-2025-A06 and CMS-CSST-2025-A08.


%





\end{document}